\documentclass{PoS}

\title{Dynamical simulation of lattice $4d$ ${\cal N}=1$ SYM%
\thanks{Work supported by: DAAD - the German Academic Exchange Service and
DFG - the German Science Foundation.}}

\ShortTitle{Dynamical simulation of lattice $4d$ ${\cal N}=1$ SYM}

\author{\speaker{K.~Demmouche}, F.~Farchioni, A.~Ferling, G.~M\"unster,
J.~Wuilloud\\
        University of M\"unster, Institute for Theoretical Physics  \\
        Wilhelm-Klemm-Strasse 9, D-48149 M\"unster, Germany\\
        E-mail: \email{k\_demm01@uni-muenster.de}}

\author{I.~Montvay\\
        Deutsches Elektronen-Synchrotron DESY, Notkestr. 85, D-22603 Hamburg, 
        Germany\\}

 \author{E.E.~Scholz\thanks{Present address: Fermi National Accelerator 
         Laboratory, Batavia, IL 60510, USA}\\
 Physics Department, Brookhaven National Laboratory, Upton, NY 11973, USA\\
}         
\abstract{The lattice provides a powerful tool to non-perturbatively investigate strongly coupled supersymmetric Yang-Mills (SYM) theories. The pure SU(2) SYM theory with one supercharge is simulated on large lattices with small Majorana gluino masses down to about $am_{\tilde g}=0.068$ with lattice spacing $a\simeq 0.125$ fm. The gluino dynamics is simulated by the Two-Step Multi-Boson (TSMB) and the Two-Step Polynomial Hybrid Monte Carlo (TS-PHMC) algorithms. Supersymmetry (SUSY) is broken explicitly by the lattice and the Wilson term and softly by the presence of a non-vanishing gluino mass. However, the recovery of SUSY is expected in the infinite volume continuum limit by tuning the bare parameters to the SUSY point in the parameter space. This scenario is studied by the determination of the low-energy mass spectrum and by means of lattice SUSY Ward-Identities (WIs).}

\FullConference{8th Conference Quark Confinement and the Hadron Spectrum\\
        September 1-6, 2008\\
        Mainz. Germany}

\begin{document}

\section{Introduction}

Non-perturbative dynamical effects like confinement and the spontaneous breaking of a discrete chiral symmetry are expected to occur in the pure ${\cal N}=1$ supersymmetric Yang-Mills (SYM) theory with SU($N_c$) as the color gauge group. The confinement is realized by colorless bound states, while the anomaly leaves only a discrete subgroup of the chiral transformations invariant, which is spontaneously broken by the non-zero gluino condensate.  Continuum low-energy effective actions \cite{Veneziano:1982ah, Farrar:1998rm} predict two chiral supermultiplets at the lower-end of the particle spectrum, each containing two scalars with opposite parities and one Majorana fermion. A massive gluino introduces a soft breaking of supersymmetry (SUSY) resolving the degeneracy in the supermultiplets.

We continue here the investigations of the DESY-M\"unster collaboration on non-perturbative features of the ${\cal N}=1$ SU(2) SYM theory using lattice simulations (for reviews, see \cite{Montvay:2001aj} and references therein.) The Wilson formulation for gluino action is applied while for the gauge (gluon) action we used pure and improved Wilson lattice actions. The Wilson formulation for the gluino action breaks both SUSY and chiral symmetry explicitly, and the presence of a non-vanishing mass term breaks SUSY softly, too. The recovery of SUSY and chiral symmetry is expected in the infinite volume continuum limit by an appropriate fine tuning of the bare gauge coupling and the bare gluino mass to their critical values. To this end, the mass spectrum of low-lying bound states of the theory is computed. Furthermore, the lattice SUSY Ward-Identities (WIs) can be checked independently, indicating that SUSY is restored in the continuum limit.%


\section{Simulation details}


\begin{table}[h]

\caption{\em Algorithmic parameters for the $\beta=2.3$ , $16^3\cdot 32$ TSMB runs. $A_{NC}$ is the acceptance rate in the noisy correction step. $\tau^{plaq}$ is the autocorrelation times of the plaquette. For $\epsilon$, $\lambda$, $n_1$, $n_2$ see text.}

\label{tab:TSMB run parameters}
\begin{center}
\begin{tabular}{c|ccccccccc}
\hline \hline
Run &$\kappa$ &\# Sweep & $A_{\rm NC}$ \%&$\tau^{plaq}$ &$\epsilon$ 
& $\lambda$ & $n_1$ & $n_2$  \\ 
\hline
$(a)$ & 0.1955 & 12500 & 50-80 & 167.6 &$2.0\cdot 10^{-5}$ & 4.0 & 40 &  800 \\
$(b)$ & 0.1960 & 23500 & 50-80 & 181.1 &$4.0\cdot 10^{-6}$ & 4.0 & 40 & 1800 \\
$(c)$ & 0.1965 & 18000 & 50-62 & 254.2 &$4.0\cdot 10^{-6}$ & 4.0 & 40 & 1800 \\
\hline \hline
\end{tabular} 
\end{center}
\end{table}

To simulate gluinos on the lattice, we used the Curci-Veneziano (CV) effective gauge action \cite{Veneziano:1982ah}. A range of lattices were simulated using the Wilson formulation for both the gluino and the gluon parts of the lattice action. In recent simulations, stout-smearing of the gauge fields used in the Dirac-Wilson operator has been applied to reduce the fluctuations of the smallest eigenvalues of the fermion matrix resulting in only a few exceptional configurations. Furthermore, in the gauge sector we used an improved lattice action, namely the tree level improved Symanzik (tlSym) gauge action (For more details see ref.~\cite{Demmouche:2008ms}).

\begin{figure}[t]
\begin{center}
\vspace{-0.5cm}
\includegraphics[width=4cm,height=7cm,angle=-90]{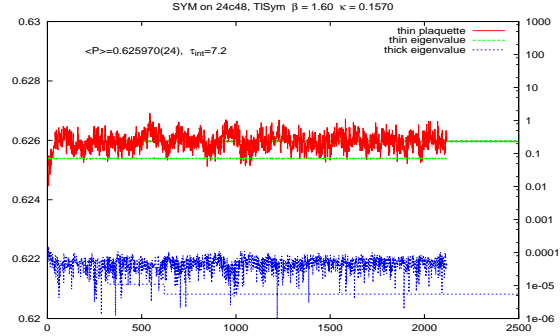}
\end{center}
\caption{History of the average plaquette and the smallest eigenvalues.}
\label{fig:history_24c48_TlSym_b160_k1570}
\end{figure}

We used two variants of dynamical updating algorithms: 1) at gauge coupling $\beta=2.3$ three ensembles at hopping parameter values $\kappa=0.1955, \kappa=0.196$ and $\kappa=0.1965$ were prepared by the Two-Step Multi-Boson (TSMB) algorithm \cite{Mont96} on $16^3\cdot 32$ lattices. 2) at gauge coupling $\beta=2.1$ several $16^3\cdot32$ and $24^3\cdot48$ lattices were generated by the Two-Step Polynomial Hybrid Monte Carlo (TS-PHMC) algorithm \cite{Montv}. The input parameters of the TSMB runs are summarized in Table \ref{tab:TSMB run parameters} and the TS-PHMC runs input parameters can be found in Table 1 of Ref.~\cite{Demmouche:2008ms}. These algorithms are based on the multi-boson representation of the fermion determinant by approximating the inverse fermion matrix by a polynomial of order $n_1$ in an interval $[\epsilon,\lambda]$ and a stochastic noisy correction (NC) step realized by a second polynomial of higher order $n_2$. The history of the average plaquette and the smallest eigenvalue of one run using stout-smearing is displayed in Fig.~\ref{fig:history_24c48_TlSym_b160_k1570}.

\section{Confinement and physical scale}

\begin{figure}[h]
\begin{center}
\includegraphics[width=6.5cm,height=4cm]{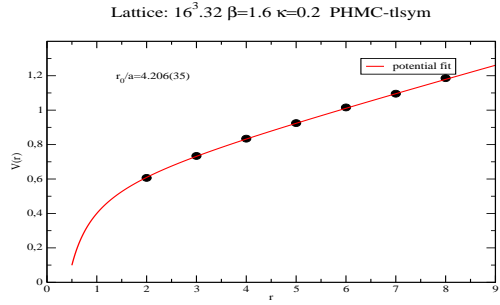}
\end{center}
\caption{The static potential.}
\label{fig:pot_phmcb16k2000Btlsym_16c}
\end{figure}

Analogy with QCD suggests that the theory presents confinement with  color singlet particle states. Confinement can be studied by computing the static quark potential $V(r)$ as a function of the spatial separation $r$ between sources in the fundamental representation of the gauge group. 
%
%
%
%
%
The basic quantities to determine the static potential are the Wilson loops. The static potential $V(r)$ in Fig.~\ref{fig:pot_phmcb16k2000Btlsym_16c} was extracted from the single-exponential fit as described in \cite{Montvay:2001aj}. We applied APE smearing $(\epsilon_{APE}=0.5,N_{APE}=4)$ for Wilson loops to improve the signal and to reduce the contamination with excited states.%

The physical scale is set by the value of the string tension or the Sommer scale parameter $r_0/a$. These are determined by two independent methods: using the force method based on the Creutz-ratios and by fitting $V(r)$ directly. At lighter gluino mass we obtained $r_0/a\simeq 8$ for TSMB runs and $r_0/a\simeq4$ for the TS-PHMC runs. Using $r_0=0.5\,{\rm fm}$, the lattice spacings are $a\approx 0.06$ fm and $a\approx 0.125$ fm for the TSMB and TS-PHMC runs, respectively. The physical size of the simulated boxes is then $L^3\simeq (1 \mbox{fm})^3$ for the TSMB sample and $L^3\simeq (2-3\,\mbox{fm})^3$ for the TS-PHMC sample.

\section{Supersymmetry and chiral limit}

\begin{figure}[h]
\begin{center}
\vspace{1.0cm}
\includegraphics[width=10.5cm,height=4cm]{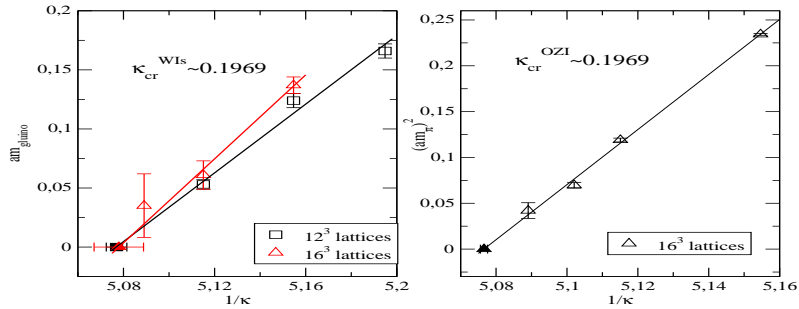}
\caption{The gluino mass (left) and the pion mass square (right) as function
of the inverse hopping parameter (TSMB runs). The data of $12^3\cdot24$ are from
Ref.~\cite{Farchioni:2004fy}}
\label{fig:wi_ozi_extrapol_TSMB}
\end{center}
\end{figure}

\begin{figure}[h]
\begin{center}
\vspace{0.3cm}
\includegraphics[width=10.5cm,height=4cm]{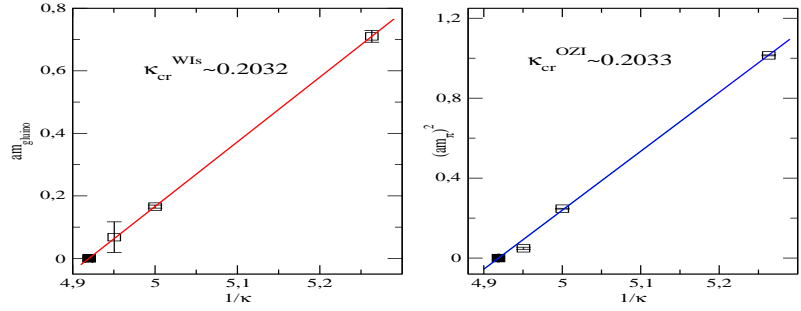}
\caption{The gluino mass (left) and the pion mass square (right) as function
of the inverse hopping parameter (TS-PHMC $16^3\cdot32$).}
\label{fig:wi_ozi_extrapol_PHMC}
\end{center}
\end{figure}

The gluino mass can be determined by means of the lattice SUSY Ward-Identities (WIs). Since it is assumed to behave linearly in $1/\kappa$, one can extrapolate the inverse hopping parameter to the massless gluino limit, where both SUSY and chiral symmetry emerge. In that way the critical value $\kappa_{\rm cr}$ of the hopping parameter can be determined. However, a second independent method based on OZI arguments \cite{Veneziano:1982ah} can be used to estimate $\kappa_{\rm cr}$, where the adjoint pion $a-\pi$ becomes massless in the chiral limit. We observed good agreement between $\kappa_{\rm cr}$ estimated from these two methods. In the case of the TSMB ensembles we found $\kappa_{\rm cr}\simeq0.1969$, and for the TS-PHMC ensembles $\kappa_{\rm cr}\simeq0.2033$. The linear extrapolations for both methods are shown in Fig.~\ref{fig:wi_ozi_extrapol_TSMB} and Fig.~\ref{fig:wi_ozi_extrapol_PHMC}.

\section{Spectrum results and conclusion}

\begin{figure}[t]
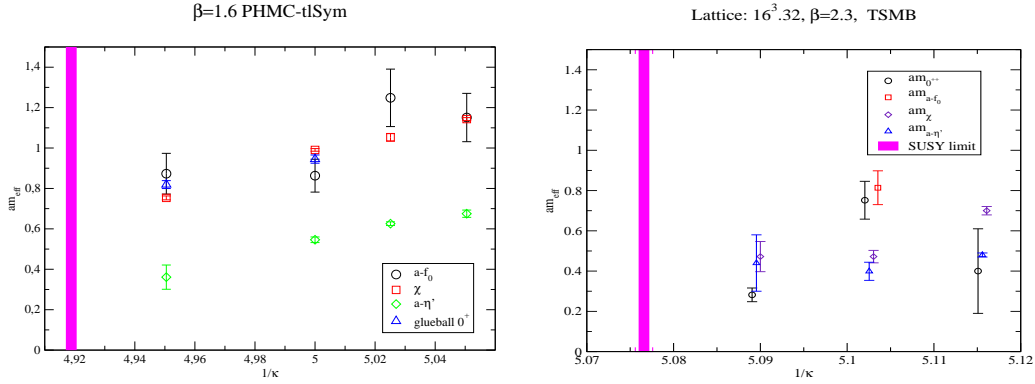

\begin{center}
\hfill 
\includegraphics[width=6.5cm,height=5cm]{Plots/spectrum_phmc.eps}\hfill
\includegraphics[width=6.5cm,height=5cm]{Plots/spectrumTSMB.eps}
\hfill$\phantom{.}$

\caption{Low-lying spectrum of ${\cal N}=1$ SU(2) SYM theory. The mass in lattice unit as function of the bare gluino mass $1/\kappa$.}
\label{fig: the mass spectrum}
\end{center}
\end{figure}

In the SUSY point the low-energy effective actions describe the lower-end particle spectrum by two massive chiral supermultiplets \cite{Veneziano:1982ah, Farrar:1998rm}. The predicted composite states are color singlet bound states: two scalars ($a-f_0$, $0^+$ glueball), two pseudoscalars ($a-\eta^\prime$, $0^-$ glueball) and two spin-$\frac12$ states $\chi$ (gluino-glueballs). 
The physical states of the two supermultiplets result in general from mixing of these states.
In lattice simulations correlation functions of interpolating fields with the corresponding quantum numbers are computed for large time separations such that the lightest state with these quantum numbers (unmixed state) dominates. By this method, the masses of the bound states are extracted. The results for the mass spectrum are presented in Fig.~\ref{fig: the mass spectrum} as a function of the inverse hopping parameter. The vertical line indicates the massless gluino region.

The results of TSMB simulations in a $(1\,{\rm fm})^3$ physical volume indicate the degeneracy of $a-\eta^\prime$ and $\chi$ masses and hence the two bound states may occur in a supermultiplet. However, in the larger volumes of TS-PHMC runs a mass difference between these two states has been observed, which appears to be larger than the expected value of the mass-splitting from the soft breaking, given by the gluino mass. We conclude that the TSMB results are dominated by large finite size effects. Furthermore, the masses of three bound states $\chi$, $a-f_0$ and $0^+$ glueball converge to one point near the region where SUSY is expected to emerge, and hence this is a signal of the heavier supermultiplet, while the $a-\eta^\prime$ remains lighter and it could belong to the lightest supermultiplet. Simulations on a finer lattice will clarify the impact on these findings of explicit SUSY breaking.

Our computations were performed on the Blue Gene L/P and JuMP at the Neumann Institute for Computing (NIC), the Cluster of the University of M\"unster and the Cluster of  RWTH Aachen.

\end{document}